\documentclass[10pt,conference]{IEEEtran}
\IEEEoverridecommandlockouts 

\usepackage[UKenglish]{babel}

\usepackage{xspace}
\usepackage{hyperref}
\usepackage[capitalise]{cleveref}
\usepackage{subcaption}
\usepackage{booktabs}
\usepackage{amssymb}
\usepackage[
  colorinlistoftodos,
  prependcaption,
  textsize=scriptsize
]{todonotes}
\usepackage{cite}
\usepackage{listings}
\usepackage{subcaption}
\usepackage{csquotes}
\usepackage{fancyvrb}

\usepackage{scratch3}
\setscratch{scale=0.45,
  else word=else,
}

\newcommand{\inlinescratch}[1]{\begin{scratch}
    #1
  \end{scratch}}
\definecolor{swamp}{HTML}{5ce581}
\newcommand{\inlinepen}[1]{\resizebox{!}{1.5ex}{\pencolor{#1}}}

\newcommand\definetool[2]{\newcommand{#1}{{\textit{#2}}\xspace}}
\newcommand\definename[2]{\newcommand{#1}{{\textit{#2}}\xspace}}
\newcommand\defineinit[2]{\newcommand{#1}{{\textsc{#2}}\xspace}}

\definetool{\Scratch}{Scratch}
\definetool{\Snap}{Snap!}
\definetool{\LitterBox}{LitterBox}
\definetool{\furball}{\textit{LitterBox\textsuperscript{+}}\!}
\definetool{\whisker}{Whisker}
\definetool{\gpt}{GPT}

\definename{\scratchblocks}{scratchblocks}
\definename{\openai}{OpenAI}
\definename{\ollama}{Ollama}

\defineinit{\llm}{llm}
\newcommand{\llms}{\textsc{llm}s\xspace}
\defineinit{\json}{json}
\defineinit{\api}{api}
\defineinit{\AST}{ast}
\defineinit{\gui}{gui}
\defineinit{\rest}{rest}
\defineinit{\id}{id}
\defineinit{\cli}{cli}

\usepackage{orcidlink}

\usepackage[final]{microtype}
\begin{document}
\title{LitterBox\textsuperscript{+}: An Extensible Framework for LLM-enhanced Scratch Static Code Analysis}

\author{\IEEEauthorblockN{Benedikt Fein \orcidlink{0000-0002-3798-845X}}
  \IEEEauthorblockA{University of Passau\\Passau, Germany}
  \and
  \IEEEauthorblockN{Florian Obermüller \orcidlink{0000-0002-6752-6205}}
  \IEEEauthorblockA{University of Passau\\Passau, Germany}
  \and
  \IEEEauthorblockN{Gordon Fraser \orcidlink{0000-0002-4364-6595}}
  \IEEEauthorblockA{University of Passau\\Passau, Germany}
}

\maketitle

\begin{abstract}
Large language models (\llms) have become an essential tool to
  support developers using traditional text-based programming
  languages, but the graphical notation of the block-based \Scratch
  programming environment inhibits the use of \llms.
To overcome this limitation, we propose the \furball framework that
  extends the \Scratch static code analysis tool \LitterBox with the
  generative abilities of \llms. By converting block-based code to a
  textual representation suitable for \llms, \furball allows users to
  query \llms about their programs, about quality issues reported by
  \LitterBox, and it allows generating code fixes. Besides offering a
  programmatic \api for these functionalities, \furball also extends
  the \Scratch user interface to make these functionalities available
  directly in the environment familiar to learners.
The framework is designed to be easily extensible with other
  prompts, \llm providers, and new features combining the program
  analysis capabilities of \LitterBox with the generative features of
  \llms.
We provide a screencast demonstrating the tool at
  \url{https://youtu.be/RZ6E0xgrIgQ}.
\end{abstract}

\begin{IEEEkeywords}
  Scratch, Block-based Programming, LLM, Automated Feedback
\end{IEEEkeywords}

\section{Introduction}

\Scratch~\cite{Maloney2010} is a popular block-based programming
environment frequently used to introduce beginners to
programming~\cite{McGill2020}. While the block-based nature simplifies
coding, beginners often lack a deep understanding of required concepts
and introduce issues into their code~\cite{Fraedrich2020}. \llms have
recently become popular to help programmers of text-based languages to
fix and improve their code, but the graphical nature of block-based
code so far hampered the application of \llms to \Scratch.

\begin{figure}[t]
  \includegraphics[width=\linewidth]{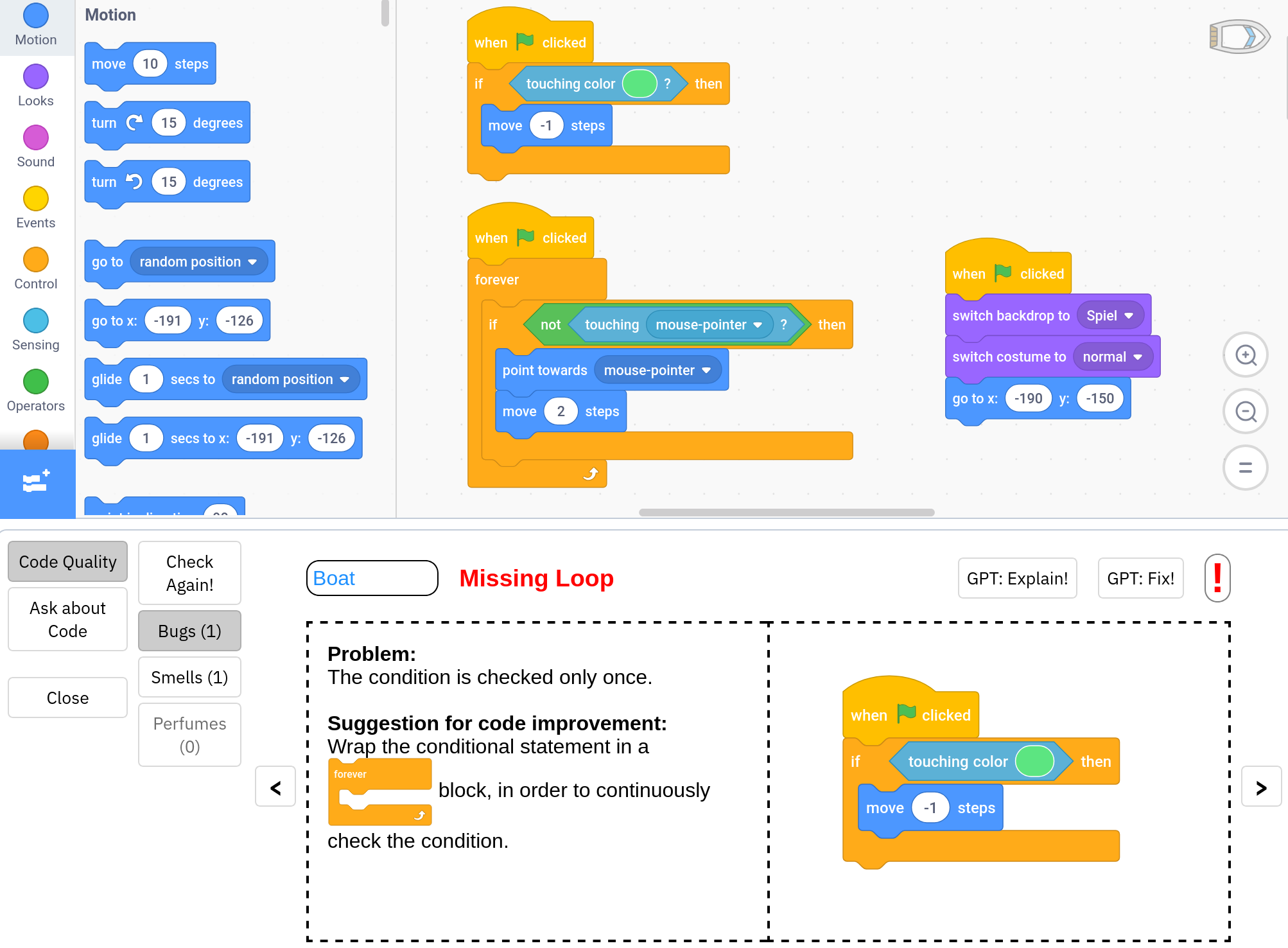}
  \caption{A buggy program displayed in the extended \Scratch-\gui, together with \llm controls to explore the bug.}\label{fig:hint-example}
\end{figure}

To bridge the gap between \Scratch programs and the promises of \llm
support, we introduce the \furball framework: It uses the \LitterBox
static code analysis framework~\cite{Fraser2021} to convert \Scratch
code to a textual representation suitable for \llms, and provides a
convenient API to query \llms about \Scratch programs. The framework
is designed to be easily extensible with different types of prompts
and queries, as well as LLM providers. It automatically provides the
context for a given query using the target code in question, offers
convenient access to common types of queries, and can build on the
existing linting features of \LitterBox to allow users to gain further
insights and even fix suggestions for the otherwise generic issue
descriptions delivered by the static analysis. In addition, \furball
extends the \Scratch user interface to enable learners to directly
make use of these features~(cf.~\cref{fig:hint-example}). This
includes a custom parser for \llm responses such that suggested fixes
can be directly applied to \Scratch projects.

\furball is also intended to be an extensible framework upon which
researchers can build creative new \Scratch{}+\llm integrations,
tapping into both the potential of \llms as well as the established
static analysis features of \LitterBox.
To support both users and researchers alike, \furball with its
command-line and Java \api{}s as well as the extended \Scratch user
interface are available as open-source code. The extension to
\LitterBox is available at
\mbox{\url{https://github.com/se2p/litterbox}} and our extensions to
the \Scratch user interface can be found at
\mbox{\url{https://github.com/se2p/NuzzleBug}}.

\section{Background}

The different shapes of \Scratch blocks determine how they can be
combined into programs, thus simplifying programming. However,
learners may still struggle to correctly apply programming concepts
and produce buggy code.  While the process of finding and fixing bugs
is an inherent part of learning to program, automated tools can offer
support.  Since \Scratch code is block-based, dedicated analysis tools
are required. For example, \LitterBox~\cite{Fraser2021} is a linter
that automatically detects bugs based on recurring patterns that are
common manifestations of misconceptions. \Cref{fig:hint-example} shows
such a bug pattern in the popular \enquote{BoatRace}
program\footnote{\url{https://scratch.mit.edu/projects/63957956/},
  accessed 2025-07-21} where a misconception caused a
\enquote{Missing Loop} bug pattern, resulting in a collision detection
only being executed once at the start of program execution. Such bug
patterns are common in \Scratch~\cite{Fraedrich2020}, and feedback on
them has been shown to support learners~\cite{Greifenstein2021}.
However, classic static analysis tools like \LitterBox can only
provide generic help, and learners have to rely on other sources for
clarifications or suggestions on how to fix their code.

While \llms are capable of providing exactly this help for text-based
languages, their application to \Scratch code is hampered by the
graphical block-based notation. Consequently, existing work on using
\llms for \Scratch tends to focus more on creative aspects, like for
example using the \llm chat to collect ideas for program features or
using the model to generate sprite
images~\cite{Chen2024,Druga2025}.
These existing approaches integrate less targeted context,
e.g.,~generic \Scratch code patterns from other
programs~\cite{Chen2024}, when students ask for support in the \llm
chat~\cite{Chen2024,Druga2025}.
In contrast, in this paper we focus on providing specific context to
the \llm in the form of the program’s code, as well as detected issues
therein to help the user to solve these issues.

Compared to prior work, we also support more extensive \llm-based
modifications of existing programs. In particular, we do not rely on
heuristic text matching semantics to collect \Scratch blocks from the
\llm{}’s output~\cite{Chen2024}, but instead developed a parser for
the widely used textual
\scratchblocks{}\footnote{\url{https://en.scratch-wiki.info/wiki/Block_Plugin/Syntax},
  accessed 2025-07-22} \Scratch code representation as part of
\furball, which allows us for example to merge fixed scripts returned
by the \llm into the abstract syntax tree of the surrounding program.

\section{User Interfaces}

The main user interface is integrated directly into the \Scratch web
interface. 
This direct integration ensures a low barrier to entry by not
requiring context switches away from the familiar programming
environment. We added a toggleable panel that shows by default issues
found by the regular \LitterBox static code analysis in the current
program~(cf.~\cref{fig:hint-example}).
For each issue, the user can request an extended \llm-generated
explanation, or request a fix for the issue, which will be
automatically integrated into the active program.
Additionally, our \gui extension offers a chat-like input where a user
can ask arbitrary questions about the current sprite or program.

While early trials of our tool showed that \llms can understand and
reason about \Scratch code, the model may still occasionally produce
misleading or wrong explanations and fixes. Since especially learners
often cannot discern such cases from helpful ones, we show a prominent
hint about the potentially wrong nature in every location where
\llm-generated content is presented. For example, the red exclamation
mark in \cref{fig:hint-example} shows this hint as tooltip when
hovered over.
Even though our tool is \llm-agnostic, we call all \llms
\enquote{\gpt{}} in the user interface, since students are likely more
familiar with \gpt as a general term for all kinds of \llms.

By requesting a different output language in the prompts, \furball can
be customised to support any natural language. While the
\scratchblocks code in the prompts remains English, the \Scratch-\gui
can still automatically display the code in the editor area in the
user’s chosen \textsc{ui} language.

For teachers or other more experienced users, we also make all
functionality of our \furball extension available via the command-line
interface~(\cli).
Like regular \LitterBox, this for example additionally allows for
batch-processing of multiple programs or automatically downloading and
processing specific programs from the official \Scratch website.
The extended \LitterBox can be started via the \cli like a regular
Java program:
\centerline{\footnotesize\texttt{java
-Dlitterbox.llm.openai.api-key=KEY}}
\centerline{\footnotesize\texttt{-jar LitterBox.jar llm --help}}
For example, to ask the \llm to analyse the \enquote{Boat} sprite of
the program and report new issues missed by \LitterBox:
\centerline{\footnotesize\texttt{[…].jar llm analyze
--path=program.sb3}}
\centerline{\footnotesize\texttt{--target=Boat --new-issues}}
All subcommands of the \cli accept the \texttt{--help} option which
explains the available subcommands and their options.

Finally, to allow future research to extend upon our framework, all
\LitterBox functionality is available via its Java \api so that it can
also be integrated as a library into other tools.
This \api allows for future extension in various directions:
(1)~Our prompt designs can be replaced by customised prompt templates.
(2)~While the user can already configure \furball to use either
externally hosted~(\openai) or self-hosted~(\ollama) models,
additional model providers could be added.\footnote{Our implementation
uses \emph{langchain4j}, \url{https://docs.langchain4j.dev/}, accessed
2025-07-14.}
(3)~New \llm-based features that make use of the existing static code
analysers and the \scratchblocks parser, or ones based on entirely new
analysers could be introduced.

For example, to change the default prompts, only a single
\enquote{prompt provider} class has to be overridden. It can then be
selected as the default prompt provider by setting a system property
when starting the \LitterBox application.
Using a similar selection mechanism, another \llm provider can be
supported by adding a new concrete implementation of our generic
\llm-provider interface to \LitterBox.
The same system properties are used by both \LitterBox
itself~(i.e.,~when using it as a command-line tool) and by our
\rest-\api extension~(i.e.,~when using it as a backend for the
\Scratch-\gui).

\section{System Architecture}\label{sec:system-arch}

\begin{figure}[t]
  \centering \includegraphics[width=\linewidth]{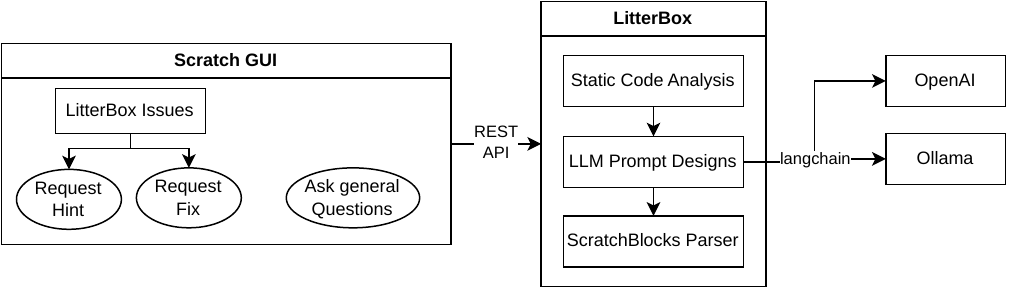}
  \caption{\furball architecture overview.}\label{fig:architecture}
\end{figure}

To support the communication between \LitterBox running as a Java
application on a server and the website in the browser, we extend
\LitterBox with a \rest-\api. When we receive a request relating to an
issue found by \LitterBox from the user interface, we combine
information from the issue itself and the program to construct a
prompt for the \llm.
From the issue itself, we extract its type~(e.g.,~Missing Loop) and
the default description added by \LitterBox.
To embed the issue in the relevant context, we also extract the source
code of the \Scratch program. To provide this code to the \llm, we
convert the program into the textual \scratchblocks format. We use the
same format as is used for example on the \Scratch community forums
and only extend it slightly to include special comments adding unique
\id{}s to the start of sprites and scripts. The \scratchblocks format
was likely seen by the \llm during training as it is common on forums
and uses a sequential structure for code, unlike \Scratch's internal
\json program representation, thus resulting in improved performance.
Unlike taking screenshots of the workspace, this removes the visual
component from the code. This avoids the problem that code blocks are
allowed to overlap in the \Scratch coding area; by using the textual
format instead of a screenshot of the code, we therefore have more
control over which part of the code is included in our prompt.

Depending on the task, we prompt the \llm to respond either with an
explanation suitable for beginner programmers or respond with a code
snippet in \scratchblocks format~(e.g.,~when asking for a fix for an
issue).
To integrate such a code response into the existing program, we add a
parser for the \scratchblocks format to \LitterBox. In case the
parsing of a script fails, we extract the script from the response and
ask the \llm to fix the syntax. If the \llm fails to repair the
\scratchblocks code, the user is informed that the \llm returned
something unuseful.
Since the \llm is instructed to retain our special \id-comments for
modified scripts, we can correlate such code snippets back to the full
original program and update the code there accordingly.

Depending on the request, we finally send the response as text, an
updated issue, or the program with the \llm-proposed fix back to the
browser where it can be displayed to the user.

\section{Features}

This section describes the three main features available via the
\Scratch-\gui and other features only available in the \cli.

\subsection{Questions About the Program}

Users might have general questions about the behaviour of their
program, or they might want to find ideas on how to continue. For this
use-case, our framework allows users to ask arbitrary questions about
the program in the \Scratch user interface.
Even though we do not use the static code analysis features in this
step, \LitterBox is still used to convert the program code into the
\scratchblocks format. We then combine the user question with this
code as relevant context to create the overall prompt for the \llm.
Depending on the user choice whether the question is specific only to
the currently selected sprite or more general about the whole program,
only the relevant code context is included in the query to the \llm.

\subsection{Explanation for LitterBox Issues}

By default, our extension shows the \LitterBox bugs and smells as
generated by the static code analysis in the user interface. Their
description is mostly generic depending on the bug type, with only
some customisation placeholders that are filled for example with the
relevant variable names.
These generic hints might make it harder for students to come up with
reproduction steps or possible fixes since they need to mentally map
the generic information to their specific program first.
The generic description lacks instructions on how to reproduce the
issue, since such steps are always program-specific.

To obtain a more detailed hint, we add a \enquote{GPT\@: Explain!}
button to the user interface~(cf.~\cref{fig:hint-example}). This sends
the issue context together with the program to \LitterBox.
The prompt includes the original issue explanation and asks the \llm
to (1)~explain how the faulty code affects the observable behaviour of
sprites and (2)~to generate a list of steps required to observe this
faulty behaviour.

Since the response is an extended explanation rather than a modified
program, it is enough to append the \llm{}’s explanation to the
original issue description and send the updated issue back to the
browser for display as addition to the previous description~(cf.~left
half of the issue display in \cref{fig:hint-example}).
In this example, \emph{GPT
4.1}\footnote{\url{https://openai.com/index/gpt-4-1/}, accessed
2025-07-21} correctly explains the issue as follows:

\vspace{1em}
\begin{Verbatim}[commandchars=\\\{\},fontfamily=helvetica,fontsize=\footnotesize]
1. How the faulty code affects observable behavior of sprites:

The static code analysis tool pointed out that a certain condition is only
checked once, instead of being checked continuously.

(a) This script will only check \emph{once}, immediately after the green flag is
clicked, if the sprite is touching the color \inlinepen{swamp}.

(b) If at that exact moment the sprite is NOT touching this color,
nothing will happen, \emph{even if the sprite touches it later}.

(c) As a result, the response (move -1 steps) won't trigger when the
sprite later comes into contact with the specified color, making the
code seem unresponsive.

(d) \emph{Observable Effect:} The Boat sprite will NOT react (move -1 steps)
when it touches the color \inlinepen{swamp} at any point after the game starts,
unless it happens to be touching it \emph{exactly} as the green flag is clicked.

2. List of steps to observe the faulty behavior:

1. \emph{Press the green flag} to start the program.

2. \emph{Move the Boat sprite} (using code or manually, depending on your
setup) so that it touches the color \inlinepen{swamp} \emph{after} the game has started.

3. \emph{Observe:} The Boat sprite does \emph{not} move backwards (-1 steps),
despite touching the color.

4. \emph{(Optional)} Restart and have the sprite start on top of \inlinepen{swamp} color. Now,
you might see it move -1 steps at game start, since the condition was
true at that moment.
\end{Verbatim}

\subsection{Automatic Issue Fixing}

To extend the support beyond providing a more specific description of
the issue, our framework can also propose fixed versions of the code
and integrate them directly into the current program.
Similar to the explanation, we added a \enquote{GPT\@: Fix!} button
into the user interface next to the shown \LitterBox issue.

When requesting the \llm to propose a suitable fix, we again prompt it
with the current issue description and the relevant code context in
\scratchblocks format.
As opposed to the previously described features, the \llm now has to
produce code in its output rather than only a natural language
description. Like for the \llm input, we rely on the \scratchblocks
format and extend \LitterBox with a suitable parser. Our parser can
process all standard \Scratch blocks, custom block definitions, and
blocks of the \Scratch \enquote{Pen} extension. In the future, its
grammar could be extended to support further \Scratch extensions.

While we observed that the \llms can in most cases produce valid
\scratchblocks code, they still often repeat similar issues, like for
example using a non-existing block \inlinescratch{\blockmove{set
rotation to \ovalnum{}}} instead of the actual
\inlinescratch{\blockmove{point in direction \ovalnum{}}} or using
braces to scope inner blocks of control statements instead of using
the correct \texttt{then} and \texttt{end} markers.
Sometimes the \llm also includes both the buggy and the fixed scripts
in its output with the script marker
comments~(cf.~\cref{sec:system-arch}) modified with an additional
\enquote{original version} or \enquote{modified version} suffix, even
though our prompts explicitly request the \llm to not modify the
special comments.
To mitigate such frequent problems, we introduced a postprocessing
step to the \llm{}’s output that applies heuristic rules fixing common
syntactical issues before passing the \scratchblocks code to the
actual parser.
In case this parsing still fails, we prompt the \llm with only the
unparseable scripts again in up to three attempts to request a
syntactically correct script instead. If the code continues to be
incorrect afterwards, we remove it and do not process it further.

To integrate the parsed scripts into existing programs, we rely on the
special comments marking the beginning of sprites and
scripts~(cf.~\cref{sec:system-arch}). Using the sprite names and block
\id{}s contained in these comments, we can align the parsed code
snippets with the original code and make the required replacements.
These replacements are performed on the internal program
representation of \LitterBox~(i.e., the abstract syntax tree), which
is substantially more flexible and reliable compared to working
directly on the textual \scratchblocks- or \json-formats.
Any sprites or scripts with unseen \id{}s are added as new code to the
program. When adding a new sprite, we use the default image of the
\Scratch cat mascot.The updated program is finally converted by \LitterBox back into the
\Scratch-native \json format and loaded into the \Scratch-\gui where
the user can inspect it in the native code editor~(upper half of
\cref{fig:hint-example}) and continue to edit it. In case the proposed
fix is not satisfactory, the user can choose to revert the change.

\subsection{Other Features}

We also implemented additional features that are currently only
available via the \cli and the Java \api.

\subsubsection{New Issue Finder}

The user can prompt the \llm to find additional issues not detected by
the \LitterBox heuristics in the program.
This can also be used to find code
\enquote{perfumes}~\cite{Obermueller2021}, i.e., parts of the code
following correct programming practices.

\subsubsection{Code Completion}

The user can select a specific script of the program by providing its
\id{} and request the \llm to extend it with suitable code.
The prompt automatically adds the surrounding sprite code as related
context to guide the \llm to provide useful code that does not
duplicate existing parts.

\section{Future Work}

Since we introduce an initial version of our extensible \furball
framework in this paper, this leaves many opportunities for future
work.
Our framework could be extended to other visual languages~(e.g.,
\Snap), but suitable linters and code converters from/into text need
to be developed first.
For use in a classroom setting where the target solution of the
program is known, additional agentic capabilities could be integrated
that automatically run \whisker~\cite{Deiner2023} or
block-based~\cite{Feldmeier2024} test suites so that either
\llm-proposed code analysis issue fixes can be checked and improved
automatically via a follow-up prompt to the \llm, or that fixes for
failing tests can be requested directly by the user.
Since the \Scratch-\gui is lacking any kind of code completion feature
at the moment, our \llm-based completion could be a valuable further
extension to the framework and its users.
Finally, all features proposed as part of \furball should be evaluated
with children to understand how such \llm-based tools are
used~\cite{Chen2024,Druga2025} to then adjust prompts and user
interface, and develop new features based on the findings.

\section{Conclusions}

In this paper we introduced our \furball framework aiming to simplify
the integration of \llms into \Scratch programming tasks. By providing
this framework as open-source, we encourage the development of new
\Scratch{}+\llm integrations going beyond the features presented in
this paper.

\section*{Acknowledgments}
\addcontentsline{toc}{section}{Acknowledgments}

We would like to thank Patric Feldmeier for his contributions to the
\scratchblocks parser implementation.  This work is supported by DFG
project FR 2955/3-3 `Pedagogy and Technology for Feedback in
Block-Based Programming' and FR 2955/5-1 `TYPES4STRINGS: Types For
Strings'.  \vfill

\bibliographystyle{IEEEtran}
\bibliography{IEEEabrv,library}

\begin{thebibliography}{10}
\providecommand{\url}[1]{#1}
\csname url@samestyle\endcsname
\providecommand{\newblock}{\relax}
\providecommand{\bibinfo}[2]{#2}
\providecommand{\BIBentrySTDinterwordspacing}{\spaceskip=0pt\relax}
\providecommand{\BIBentryALTinterwordstretchfactor}{4}
\providecommand{\BIBentryALTinterwordspacing}{\spaceskip=\fontdimen2\font plus
\BIBentryALTinterwordstretchfactor\fontdimen3\font minus
  \fontdimen4\font\relax}
\providecommand{\BIBforeignlanguage}[2]{{%
\expandafter\ifx\csname l@#1\endcsname\relax
\typeout{** WARNING: IEEEtran.bst: No hyphenation pattern has been}%
\typeout{** loaded for the language `#1'. Using the pattern for}%
\typeout{** the default language instead.}%
\else
\language=\csname l@#1\endcsname
\fi
#2}}
\providecommand{\BIBdecl}{\relax}
\BIBdecl

\bibitem{Maloney2010}
J.~Maloney, M.~Resnick, N.~Rusk, B.~Silverman, and E.~Eastmond, ``The {S}cratch
  programming language and environment,'' \emph{ACM Transactions on Computing
  Education (TOCE)}, vol.~10, no.~4, Nov. 2010.

\bibitem{McGill2020}
M.~M. McGill and A.~Decker, ``Tools, languages, and environments used in
  primary and secondary computing education,'' in \emph{Conf. on Innovation and
  Technology in Computer Science Education}.\hskip 1em plus 0.5em minus
  0.4em\relax {ACM}, 2020.

\bibitem{Fraedrich2020}
C.~Fr\"{a}drich, F.~Oberm\"{u}ller, N.~K\"{o}rber, U.~Heuer, and G.~Fraser,
  ``Common bugs in {S}cratch programs,'' in \emph{Conference on Innovation and
  Technology in Computer Science Education~(ITiCSE)}.\hskip 1em plus 0.5em
  minus 0.4em\relax ACM, 2020.

\bibitem{Fraser2021}
G.~Fraser, U.~Heuer, N.~K{\"o}rber, F.~Oberm{\"u}ller, and E.~Wasmeier,
  ``Litterbox: {A} linter for {S}cratch programs,'' in \emph{ICSE-SEET}.\hskip
  1em plus 0.5em minus 0.4em\relax IEEE, 2021.

\bibitem{Greifenstein2021}
L.~Greifenstein, F.~Oberm{\"u}ller, E.~Wasmeier, U.~Heuer, and G.~Fraser,
  ``Effects of hints on debugging {S}cratch programs: An empirical study with
  primary school teachers in training,'' in \emph{WiPSCE}.\hskip 1em plus 0.5em
  minus 0.4em\relax {ACM}, Oct. 2021.

\bibitem{Chen2024}
L.~Chen, S.~Xiao, Y.~Chen, Y.~Song, R.~Wu, and L.~Sun, ``{C}hat{S}cratch: An
  {AI}-augmented system toward autonomous visual programming learning for
  children aged 6-12,'' in \emph{CHI Conference on Human Factors in Computing
  Systems}.\hskip 1em plus 0.5em minus 0.4em\relax {ACM}, May 2024.

\bibitem{Druga2025}
S.~Druga and A.~J. Ko, ``{S}cratch {C}opilot: Supporting youth creative coding
  with {AI},'' in \emph{Interaction Design and Children}.\hskip 1em plus 0.5em
  minus 0.4em\relax {ACM}, Jun. 2025.

\bibitem{Obermueller2021}
F.~Obermüller, L.~Bloch, L.~Greifenstein, U.~Heuer, and G.~Fraser, ``Code
  perfumes: Reporting good code to encourage learners,'' in
  \emph{WiPSCE}.\hskip 1em plus 0.5em minus 0.4em\relax {ACM}, Oct. 2021.

\bibitem{Deiner2023}
A.~Deiner, P.~Feldmeier, G.~Fraser, S.~Schweikl, and W.~Wang, ``Automated test
  generation for {S}cratch programs,'' \emph{Empirical Software Engineering},
  vol.~28, no.~3, May 2023.

\bibitem{Feldmeier2024}
P.~Feldmeier, G.~Fraser, U.~Heuer, F.~Obermüller, and S.~Steckenbiller, ``A
  block-based testing framework for {S}cratch,'' in \emph{Koli Calling}.\hskip
  1em plus 0.5em minus 0.4em\relax ACM, Nov. 2024.

\end{thebibliography}

\end{document}